\documentclass[onecolumn, groupedaddress, showkeys]{revtex4}
\usepackage{setspace}
\usepackage[dvips]{color}
\usepackage{xcolor}
\usepackage{amsfonts}
\usepackage{amsmath}
\usepackage{amssymb}
\usepackage{graphicx}
\usepackage{mathrsfs}
\usepackage{bbm}
\usepackage{bm}
\usepackage{multirow}

\begin{document}

\newcommand{\Secref}[1]{Sec.~\ref{#1}}
\newcommand{\dd}{d}%\textrm{d}
\newcommand{\pd}{\partial}
\newcommand{\myU}{\mathcal{U}}
\newcommand{\myr}{q}
\newcommand{\Urho}{U_{\rho}}
\newcommand{\myalpha}{\alpha_*}
\newcommand{\bd}[1]{\mathbf{#1}}
\newcommand{\Eq}[1]{Eq.~(\ref{#1})}
\newcommand{\Eqn}[1]{Eq.~(\ref{#1})}
\newcommand{\Eqns}[1]{Eqns.~(\ref{#1})}
\newcommand{\Figref}[1]{Fig.~\ref{#1}}
\newtheorem{theorem}{Theorem}
\newcommand{\me}{\textrm{m}_{\textrm{e}}}
\newcommand{\sgn}{\textrm{sign}}
\newcommand*{\bfrac}[2]{\genfrac{\lbrace}{\rbrace}{0pt}{}{#1}{#2}}

\renewcommand{\thesection}{\arabic{section}}
\renewcommand{\thesubsection}{\thesection.\arabic{subsection}}
\renewcommand{\thesubsubsection}{\thesubsection.\arabic{subsubsection}}

%\title{Dark Energy, Cosmological Units and Masses of Photons and Neutrinos}
% Force line breaks with \\ 

\title{Cosmological Mass of the Photon and Dark Energy as its Bose-Einstein Condensate in de Sitter Space}

\author{Lorenzo Gallerani Resca} 
\email{resca@cua.edu}
\homepage{http://physics.cua.edu/people/faculty/homepage.cfm} 

\affiliation{Department of Physics and Vitreous State Laboratory, 
The Catholic University of America,  
Washington, DC 20064}

\date{\today} 

\begin{abstract}

\textbf{ABSTRACT} --- I develop a physical picture of dark energy (DE) based on fundamental principles and constants of quantum mechanics (QM) and general relativity (GR) theories. It derives from a conjecture of non-zero masses for nearly standard-model photons or gluons, based on QM localization at a cosmological scale. Dark energy is associated with de Sitter space and that has a fundamentally invariant event horizon, which provides the basis for my DE model. I conceive of DE as a Bose-Einstein condensate (BEC) of cosmologically massive photons and I estimate fundamentally the binding energy per particle originating from an effectively attractive QM potential in that BEC. Since massive photons may stand at rest in a de Sitter universe with flat spatial geometry, I solve the time-independent Schr\"{o}dinger equation for a non-relativistic attractive spherical-well potential self-confining at the de Sitter horizon. The minimal critical potential depth that binds a particle state at the top of that well, combined with the prototypical condition of dark energy-pressure relation in the standard flat $\Lambda$-CDM model, provides an estimate of the photon mass, $m_g$. That is supported by an independent calculation of the vacuum energy of the BEC in a de Sitter static metric with coordinate-time slicing. I also consider classical gravitational collapse for a uniform dark energy density, approaching Schwarzschild condition. These QM and GR estimates provide compatible accounts of dark energy condensation, bridging a chasm between nuclear and cosmological scales. I then investigate statistical properties of equilibrium between the $g$-BEC phase and the ordinary `vapor' phase of $m_g$ photons. Resulting corrections to the Planck spectrum of the CMB are too small to be detectable, at least currently. Most notably, I consider a system of cosmological units, or `$g$-units,' that complements the fundamental system of Planck units in various ways. The geometric mean of Planck and $g$-mass turns out to be remarkably close to current estimates of neutrino masses, suggesting that even masses of the lightest known fermions may be deeply related to both GR and QM fundamental constants $\Lambda$, $G$, $c$ and $h$.

\end{abstract}

\keywords{photon mass; neutrino mass; cosmological constant; cosmological units; vacuum energy; dark energy; dark matter; Bose-Einstein condensate; general theory of relativity. --- \textbf{EMAIL}: resca@cua.edu}  

%Use showkeys class option if keyword display desired

\maketitle

\section{Introduction}\label{Introduction} 

In a previous paper I proposed non-zero bare masses for photons and gluons, derived from quantum-mechanical localization at a cosmological scale.\cite{RescaPhoton} My basic assumption is that            

\begin{equation}\label{fundamental} 
\lambda_g = \frac{h}{m_g c} = C_g^{-1} L_g = C_g^{-1} \sqrt{\frac{1}{\Lambda}} \simeq C_g^{-1} \mathrm{x} 10^{10} \mathrm{ly} .
\end{equation}  
In \Eq{fundamental}, $\Lambda$ is Einstein's cosmological constant, while $\lambda_g$ is the Compton wavelength of originally massless gauge bosons in the standard model (SM) of elementary particles. Those become endowed with a minimal but finite mass      
\begin{equation}\label{mass}
m_g c^2 = \frac{h c}{\lambda_g} = C_g M_g c^2 = C_g h c \sqrt{\Lambda} \simeq C_g (1.4 \mathrm{x} 10^{-41}) m_p c^2 ,
\end{equation}  
with a value given here in units of the proton rest-energy mass, $m_p c^2 \simeq 938 \mathrm{MeV}$. 

Beyond my original conjecture, I introduce in \Eq{fundamental} a numerical coefficient, $C_g$, which I estimate or constrain for photons by two alternative approaches. One approach involves a non-relativistic estimate of the effective binding energy for the Bose-Einstein condensate (BEC) of $m_g$-photons confined within de Sitter horizon. An alternative approach is based on the vacuum or zero-point energy of the $g$-BEC in de Sitter static metric. Remarkably, that vacuum energy can match dark energy (DE) within a single order of magnitude.

My purpose is to formulate a basic physical picture of dark energy and pressure that may be reasonably consistent with the standard flat $\Lambda$-CDM (cold dark matter) model of hot big bang cosmology and most fundamental principles of quantum mechanics (QM) and general relativity (GR) theories.

A most notable outcome of my formulation is the introduction of cosmological units $L_g$, $M_g$ and $T_g$, derived from the universal constants $\Lambda$, $c$ and $h$. These `$g$-units' complement Plank units, $l_P$, $m_P$ and $t_P$, derived from $G$, $c$ and $h$, in various ways and corresponding uncertainty relations. The fundamental ratio between corresponding Planck and $g$-units (of mass or inverse length) is approximately $2.4 \mathrm{x} 10^{60}$, reflecting the vastness between microscopic and cosmological scales of the `observable' universe. Yet the geometric mean of the `maximal' Planck mass and the `minimal' $g$-mass, which falls in the logarithmic middle of that vast range, stands withing a single order of magnitude of current estimates of neutrino masses. Chances that this is yet another mere coincidence seem remote. I thus suppose that even masses of the lightest known fermions are related to basic combinations of QM and GR fundamental constants. 

Namely, I discuss in this paper that neutrino masses may be related to dark-energy density as

\begin{align}\label{geometric neutrino}
m_{\nu} \simeq \sqrt {M_g m_P} = \bigg [ \frac{h^3 \Lambda}{c G} \bigg ]^{1/4}
=  c^{-2} \bigg [ 8 \pi (h c)^3 \rho_{\Lambda} \bigg ]^{1/4}  \simeq  2 \mathrm{x} 10^{-2} e V / c^2 .
\end{align}    

Major developments of observational and theoretical dark energy (DE) and dark matter (DM) research continue to challenge the boundaries of the standard $\Lambda$-CDM model and other fundamental concepts of modern cosmology.\cite{APS-CMB, astrobites-CMB, Riess, Prat, Capozziello, Raveri, Jones, abdalla2022cosmology, Bull, Durrer, Steinhardt, Dadhich, BianchiRovelli, Weinberg, wikizero, Bertone, Harko, Harko-Bohmer, boddy2022astrophysical} For the basic perspective of this paper, I may assume that the standard $\Lambda$-CDM model holds with sufficient accuracy. Thus I maintain that, beside Einstein's constant, $8 \pi G/c^4$, only the cosmological term can be minimally introduced in Einstein field equations without upsetting neither their stress-energy tensor conservation laws nor their general coordinate invariance, requiring a second universal and fundamental constant, $\Lambda$, with dimensions of an inverse square length.      

The extent to which my basic description of DE and my estimates of photon and neutrino masses may be confirmed and contribute to far reaching endeavors remains to be seen. In this paper I may only provide estimates that follow from my basic assumptions and evaluate whether corresponding corrections may be testable at present. Nevertheless, if correct, developments of my conception may be bound to have a major impact on fundamental physical theories and observations.

Currently, there are advanced theories that develop techniques potentially applicable to my basic cosmological description of dark energy and dark matter as well. Those include studies of ultra-light bosonic scalar and vector field dark matter,\cite{Calmet, ChavanisMass, Chavanis, Hwang, Hui, Hu, Li, Siemonsen, Tsukada, Morikawa, Nishiyama, Ferreira, LeeA} and also theories of gravitational-vacuum and dark-energy stars.\cite{Mazur, Chapline, Laughlin, VisserGrava}
%and repulsive gravity regions in gravitational collapse.\cite{Giambo} 

This paper is developed at a basic level, designed to be fully accessible to general readers. Technical issues are kept to a minimum or mainly referenced to standard literature.\cite{Mandl, Ballentine, MTW, Schutz2Ed, Hobson, Rindler, Narlikar, SchutzGM, Wheeler, Hartle, Wald, Visser, Pathria} 

%Though most experts probably know and understand all that much better than I do, I do not apologize for briefly reviewing such most beautiful ideas in the history of human thought, although many others have done so much better before. 
Thus my paper is structured as follows. 

In \Secref{FLRW}, \Secref{de Sitter} and \Secref{Perfect fluids}, I review the essential elements of the standard framework of hot big bang cosmology within which I am bound to frame my basic formulation of dark energy structure.   

That is developed in \Secref{DarkEnergy}, \Secref{Flat-space} and \Secref{Zero-point}. Most remarkably in the latter, I can identify the dark energy density, $\rho_{\Lambda}$, with my zero-point energy density of the BEC, $\rho'_{0g}$, within a single order of magnitude. By comparison, standard quantum field theory (QFT) calculations overestimate vacuum energy by as many as 120 orders of magnitude.\cite{Bull, Durrer, Steinhardt, Dadhich, BianchiRovelli, Weinberg, wikizero, Mandl, Ballentine, Hobson, Visser} In fact, the ratio of the Casimir electromagnetic vacuum energy density, $\rho_{CEM}$, to my $\rho'_{0g}$ is of the order of $8 \pi^2 (L_g/l_P)^2 \simeq 4.55 \mathrm{x} 10^{122}$.

Comparisons between my estimates of dark energy condensation and a classical GR perspective on self-gravitation and collapse are drawn in \Secref{DarkPressure}.

In \Secref{CMB}, I investigate statistical properties of equilibrium between the $g$-BEC phase and the ordinary vapor phase of cosmologically massive photons. Relations and comparisons with the Planck spectrum and measurements of the cosmic microwave background (CMB) are then evaluated and discussed.

In \Secref{GPP}, I refer to advanced theories of Gross-Pitaevskii-Poisson systems and their relativistic generalizations of hydrodynamic equations that may be applicable to my formulation of cosmologically massive $g$-photons as well.

In \Secref{Time}, I describe possible relations between QM `universal time' and GR `cosmological time.'

In \Secref{Units}, I introduce cosmological units that complement Planck units in opposite limits. A remarkable outcome of that formulation, derived in \Secref{Neutrino}, is that the geometric mean between the `minimal' and the `maximal' mass turns out to be tantalizingly close to observational values of neutrino masses: \Eq{geometric neutrino}.

Some more technical and speculative considerations about gluons and gravitons, but relevant nonetheless, are summarized in \Secref{gluons}. 

In \Secref{Conclusions}, I draw the main conclusions of my work. 

Last but not least, in the Appendix of \Secref{Appendix}, I refer to alternative work that may lead to remarkable developments and new perspectives.

\section{Einstein field equations and FLRW geometry}\label{FLRW}

My conjecture is framed in the context of the cosmological principle and observation that our universe is homogeneous and isotropic on a very large scale,\cite{MTW} which lead to a relatively simple Robertson-Walker (RW) metric. That is derived, for example, on p. 343 of Ref. \onlinecite{Schutz2Ed} as 

\begin{align}\label{FLRWmetric}
ds^2  =  & g_{\mu \nu} dx^{\mu} dx^{\nu} =  -(c dt)^2 + R^2(t) \bigg[ \frac{(dr)^2}{1 - \kappa r^2} +   r^2 \bigg( (d \theta)^2 + \sin^2 \theta (d \phi)^2 \bigg) \bigg] .
\end{align}
Flat, closed, open universes correspond to `spatial curvatures' $\kappa = 0, +1, -1$, respectively.\cite{Hobson}

Einstein's field equations may be expressed as

\begin{equation}\label{Einstein}
G^{\mu \nu} = - \Lambda g^{\mu \nu} + \frac{8 \pi G} {c^4} T^{\mu \nu} . 
\end{equation}  

One may assume that the stress-energy tensor originates from a material `perfect fluid' with pressure, $p$, and energy density, $\rho$, given by

\begin{equation}\label{PerfectFluid}
T^{\mu \nu} =  p g^{\mu \nu} + (\rho + p) U^{\mu} U^{\nu} / c^2 , 
\end{equation}  
where $U^{\mu} = dx^{\mu} / dt$ denotes the four-velocity of the fluid element. 

One may further suppose that the cosmological term also originates from a peculiar perfect fluid defined as having

\begin{equation}\label{PerfectFluidLambda}
T_{\Lambda}^{\mu \nu} = \frac{c^4} {8 \pi G} ( - \Lambda g^{\mu \nu} ) = p_{\Lambda} g^{\mu \nu} + (\rho_{\Lambda} + p_{\Lambda}) U^{\mu} U^{\nu} / c^2 . 
\end{equation}  
This definition is based on the assumption that
\begin{equation}\label{darkenergy}
\rho_{\Lambda} = - p_{\Lambda} = \frac{c^{4} \Lambda}{8 \pi G} \simeq 3.6 m_p c^2 / m^3 \simeq  5.3 \mathrm{x} 10^{-15} \mathrm{atm} .
\end{equation}  

Strict equality between a negative pressure or `tension' and energy density provides a basis for the standard $\Lambda$-CDM model. Notwithstanding its wide range of successes,\cite{Schutz2Ed, Hobson} some discrepancies by more than 5\%, or above three standard deviations, mainly between supernova measurements and Planck spacecraft CMB data, keep challenging the applicability of the $\Lambda$-CDM model.\cite{APS-CMB, astrobites-CMB, Riess, Prat, Capozziello, Raveri, Jones, abdalla2022cosmology} Thus it remains to be seen to what extent the ratio of $p_{\Lambda} / \rho_{\Lambda}$ may or may not equal $-1$ in particular.\cite{Steinhardt}

Resolving these issues is undoubtedly bound have most consequential theoretical and observational implications.\cite{Prat, Capozziello, Raveri, Jones, abdalla2022cosmology, Bull, Durrer, Steinhardt, Dadhich, BianchiRovelli, Weinberg, wikizero, Bertone, Harko, Harko-Bohmer, boddy2022astrophysical}. Currently, however, that lies beyond the scope of estimates that I provide in this paper. Thus, let me assume that \Eq{darkenergy} holds more or less accurately for my most immediate concerns.\cite{Jones}         

Mathematically, the RW metric and its simplicity derive from application of Weyl's postulate and the cosmological principle to a maximally symmetric space.\cite{Hobson, Rindler, Narlikar, SchutzGM} That applies to theories even outside GR. Within GR, application of Einstein field equations to the RW metric leads to two Friedmann-Lema\^{i}tre (FL) equations that determine the dynamical evolution of the universe `scale factor,' $R(t)$, as a function of `cosmic time,' $t$. I refer to that metric evolution as FLRW geometry or geometrodynamics.\cite{MTW, Wheeler} Derivations of FL equations are generally provided: see, for instance, p. 355 of Ref. \onlinecite{Schutz2Ed} or p. 379 of Ref. \onlinecite{Hobson}.

The first FL equation derives from the $tt$-component of \Eq{Einstein} and may be cast as
\begin{align}\label{firstFL}
( \dot{R})^2 = & - c^2 \kappa - W(R) =  - c^2 \kappa - \bigg[ -\frac{8 \pi G} {3 c^2} (\rho R^2) - \frac{c^2} {3} (\Lambda R^2) \bigg] .
\end{align}
Differentiation with respect to time is generally denoted by a `dot' over the variable, e.g., $ \dot{R} \equiv dR / dt$.

Formally, \Eq{firstFL} may be interpreted as conserving the `total energy,' $- c^2 \kappa$, of a `kinetic' term, $ ( \dot{R})^2 $, plus an `effective potential' term, $W(R)$. That $W(R)$ is generally `hill-shaped' and negative for $\Lambda > 0$. At relatively small $R$, the first term, originating from the matter and radiation energy density, $\rho$, dominates $W(R)$, attractively. At relatively large $R$, the second term, originating from dark energy density, $\rho_{\Lambda}$, dominates $W(R)$, repulsively. At the maximum of $W(R)$, or least negative, these two terms equilibrate. Should that maximum exactly coincide with $- c^2 \kappa$ for $\kappa = +1$, a static universe would result, as originally conceived by Einstein (1917). Finally it was noted by Eddington (1930) that such a static universe is hopelessly unstable: see p. 759 of Ref. \onlinecite{MTW} and p. 408 of Ref. \onlinecite{Hobson}. There is of course a lot more to that story.\cite{BianchiRovelli, Weinberg} Yet, reports of Einstein's blunders have been exaggerated. 

Compared to the enormous complexity of Einstein's \Eq{Einstein} from which it was derived, the FL first-order ordinary differential \Eq{firstFL} is remarkably simple, ultimately as a consequence of the maximal symmetry assumed in the RW metric, \Eq{FLRWmetric}. Yet \Eq{firstFL} fundamentally rules the geometrodynamics of all the observable universe on its largest scale!
 
A second and independent FL equation derives from a combination of $rr$- and $tt$-components of \Eq{Einstein} and may be cast as      
\begin{equation}\label{secondFL}
\ddot{R} = - \frac{4 \pi G} {3 c^2} (\rho + 3 p ) R + \frac{c^2} {3} \Lambda R .
\end{equation}
Both FL equations can be uniquely solved when an equation of state between pressure and energy density, $p = p(\rho)$, is specifically provided.

After a `hot big bang,' energy densities of radiation and matter dominated a decelerating expansion of $R(t)$ in the early universe. Dark energy did not affect that until much later, when it turned that expansion into acceleration around $ 7 \mathrm{x} 10^{9}$ years. Indeed, energy densities of radiation and matter decrease as $R^{-4}$ and $R^{-3}$, respectively, whereas the dark energy density remains constant and independent of $R$.           

In terms of cosmic time, $t$, the universe scale factor becomes asymptotically

\begin{equation}\label{deSitterExpansion}
R(t) =  \tilde{R} e^{c t \sqrt{\Lambda /3}}  
%R(t) =  \tilde{R} \mathrm{exp} (c t \sqrt{\Lambda /3})
\end{equation}  
on account of overwhelming dark energy. That expansion also creates an `event horizon' (EH) that ultimately reduces to a constant

\begin{equation}\label{horizon}
a_{\Lambda} = \sqrt{3/ \Lambda} 
\end{equation}  
for all future times.\cite{Hobson, Rindler, Narlikar} The Hubble parameter, $H(t) \equiv \dot{R}/R$, also approaches asymptotically a constant $H_{\Lambda} = c / a_{\Lambda}$. In emptied space-time, with $T^{\mu \nu} = 0$, these results are exact for $\kappa = 0$, while they become asymptotically correct for $t >> a_{\Lambda}/c$ if $\kappa = \pm{1}$.  

Density parameters, $\Omega_i (t)$, are defined by dividing all terms in FL \Eq{firstFL} by $(\dot{R})^2$. Following the $R(t)$ expansion, dark energy density currently amounts already to about $\Omega_{\Lambda, 0} =  c^2 \Lambda [H(t_0)]^{-2} / 3 \simeq 0.7$, and it is bound to further overwhelm exponentially all other density parameters, as $H(t)$ will further reduce to $H_{\Lambda}$ in the future.

That certainly includes the spatial curvature parameter, $\Omega_{\kappa} (t) = - c^2 \kappa / \dot{R}^2$, which raises nonetheless a fundamental question regarding the so called \textit{flatness problem}. First of all, if $\kappa = \pm{1}$, FL \Eq{firstFL} is not manifestly independent of an arbitrary factor, $\tilde{R}$, that should generally pertain to the universe scale factor, $R(t)$. More importantly, current cosmological observations indicate that, if $\kappa = \pm{1}$, the magnitude of $\Omega_{\kappa} (t)$ had to be finely tuned to a very small value at early epochs, grow exponentially to still a rather small value at the deceleration/acceleration change-over around $ 7 \mathrm{x} 10^{9}$ years, and then `refocus' back to a small value at the present time. 

Rather, data from supernovae, cosmic microwave background (CMB), and studies of evolutions of galaxy clusters suggest that $\kappa = 0$. The power spectrum of matter-density perturbations in the CMB and the angular position and scale of the first acoustic peak provide further support for a flat $\Lambda$-CDM model.\cite{Schutz2Ed, Hobson} That mainly underlies my model of the dark energy $g$-BEC in \Secref{DarkEnergy}.  

\section{de Sitter space}\label{de Sitter}

In $n = 4$ space-time dimensions, de Sitter space, $dS_4$, is the maximally symmetric Lorentzian manifold with constant positive curvature that can be embedded in a 5-dimensional Minkowski space-time, $M(1,4)$. It is a solution of Einstein's \Eq{Einstein} with $T^{\mu \nu} = 0$. Thus, $dS_4$ is a 4-dimensional `Einstein manifold,' with Ricci tensor $R^{\mu \nu} = \Lambda g^{\mu \nu}$ and a positive constant Ricci scalar curvature, $g_{\mu \nu} R^{\mu \nu} = 4 \Lambda > 0$, throughout the manifold.\cite{Rindler, Narlikar, Moschella, Ibison, IbisonTwo} That corresponds to a subset of FLRW geometry.      

Introducing particular definitions of a time-like coordinate, $x^0$, and a space-like coordinate, $x^4$, in the $M(1,4)$ ambient space, metrics and coordinates of $dS_4$ can be related to those of FLRW geometry. As shown by Lema\^{i}tre in 1925, \textit{flat slicing} of $dS_4$ corresponds to setting $\kappa = 0$ in \Eq{FLRWmetric}, with \Eq{deSitterExpansion} satisfied exactly for any arbitrary constant, $\tilde{R}$. \textit{Closed} or \textit{open slicing} of $dS_4$ correspond to setting in \Eq{FLRWmetric} $\kappa = +1$ and $R(t) = a_{\Lambda} \mathrm{cosh} (c t / a_{\Lambda})$, or $\kappa = -1$ and $R(t) = a_{\Lambda} \mathrm{sinh} (c t / a_{\Lambda})$, respectively. 

Other choices of $dS_4$ slicing and coordinates are possible. Most notably, \textit{static coordinates} yield the metric

\begin{align}\label{staticmetric}
ds^2  =  & g_{\mu' \nu'} dx^{\mu'} dx^{\nu'} =  -\bigg( 1- \frac{r'^2}{a_{\Lambda}^2} \bigg)(c dt')^2 + \bigg(1 - \frac{r'^2}{a_{\Lambda}^2} \bigg)^{-1} (dr')^2 +  r'^2 \bigg( (d \theta)^2 + \sin^2 \theta (d \phi)^2 \bigg).
\end{align}
I will refer to \Eq{staticmetric} as de Sitter's static metric, although de Sitter (1917) originally introduced other coordinates.\cite{Weinberg} De Sitter was influenced by Einstein's cosmological idea of a static closed universe. In fact, de Sitter's metric in \Eq{staticmetric} has closed spatial sections, corresponding to $\kappa = +1$ in FLRW geometry.\cite{Moschella, Ibison, IbisonTwo}

De Sitter's metric is `static' in the sense that its $g_{\mu' \nu'}$ metric tensor components are time-independent and the line element is invariant under time-reversal.\cite{Schutz2Ed} It looks similar to Schwarzschild metric, in the sense that we can cast both metrics in the form

\begin{align}\label{comparison}
ds^2  = -\bigg( 1 + \frac{2}{c^2} V(r') \bigg)(c dt')^2 + \bigg(1 + \frac{2}{c^2} V(r') \bigg)^{-1} (dr')^2 +
       r'^2 \bigg( (d \theta)^2 + \sin^2 \theta (d \phi)^2 \bigg).
\end{align}
For Schwarzschild metric, $V_S(r') = - G M / r'$ corresponds to the familiar Newtonian \textit{attractive gravitational} potential. For de Sitter metric, $V_{dS}(r') = - ({c^2} / 2) ({r'^2} / {a_{\Lambda}^2})$ corresponds to a \textit{quadratic repulsive} potential with a constant \textit{anti-gravitational acceleration}, consistent with our understanding of dark energy. 

For both Schwarzschild and de Sitter metrics, `time slicing' generates an `atemporal space.' Notice, however, that the usual interpretation of time-like and space-like coordinates holds in the \textit{interior} of the $a_{\Lambda}$ horizon in \Eq{staticmetric}, but in the \textit{exterior} of the Schwarzschild radius, $R'_S = 2 M G / c^2$. Furthermore, neither metric is fundamentally static, as more deeply understood in terms of other coordinates.\cite{Price, PricePRD, Resca, Rafael, Wormhole}

Laboriously,\cite{Ibison, IbisonTwo} one can find the transformation between $dS_4$ static coordinates and those corresponding to the FLRW geometry with $\kappa = +1$. In my formalism, that transformation is
\begin{equation}\label{TransformationOne}
r' = R(t) r = a_{\Lambda} \mathrm{cosh} (c t / a_{\Lambda}) r ,
\end{equation}  
and
\begin{equation}\label{TransformationTwo}
\mathrm{tanh} (c t' / a_{\Lambda}) = (1 - r^2)^{-1/2} \mathrm{tanh} (c t / a_{\Lambda}) .
\end{equation}  

In this context, I have not reported explicit expressions of event horizons. Suffice it to say that for $dS_4$ the EH is an invariant constant, $a_{\Lambda}$, for any choice of coordinates and any time-slicing.\cite{Hobson, Rindler, Narlikar, Moschella, Ibison, IbisonTwo} In FLRW coordinates, that can be expressed as $a_{\Lambda} = R(t) r_t^{(EH)}$, where $t$ is any current cosmic time. Thus, \Eq{TransformationOne} applies to the curvature coordinate, $r$, of FLRW geometry with $\kappa = +1$ as long as $r < r_t^{(EH)} = 1 / \mathrm{cosh} (c t / a_{\Lambda})$ and $r' < a_{\Lambda}$ remain consistently within the EH.   

It is my understanding that there is a bijection only between $dS_4$ static coordinates and FLRW coordinates with $\kappa = +1$. If there were other bijections between $dS_4$ static coordinates and FLRW coordinates with either $\kappa = 0$ or $\kappa = -1$, then there would also be bijections between FLRW coordinates with any $\kappa$, which would violate a basic tenet of FLRW geometry. On the other hand, Sec. II of Ref. \onlinecite{Weinberg} may suggest that de Sitter's original static coordinates are compatible with FLRW coordinates with $\kappa = 0$. As a matter of fact, that would strengthen the consistency of later assumptions of mine, particularly when comparing the results of \Secref{Flat-space} and \Secref{Zero-point}. Technically, however, application of de Sitter's \textit{original} static coordinates to my formulation is more cumbersome and will not be pursued in this paper. In any case, the assumption that dark energy is tied to $dS_4$ and the fact that $dS_4$ intrinsically has a robust event horizon for any coordinate system and whatever value of $\kappa$ is what matters most to support my following conjecture and model of dark energy and pressure. 

Using techniques and notations previously developed,\cite{Resca, Rafael, Wormhole, Cortical} it is instructive to determine geodesic equations and orbits in $dS_4$, using static coordinates in particular. For a particle with $m > 0$, I obtain
\begin{equation}\label{geodesicradial}
   \bigg(\frac{d r'}{d\tau}\bigg)^2
    = c^2 ( \tilde{E}^2 - 1) + c^2 \frac{r'^2}{a_{\Lambda}^2} - \bigg (1 - \frac{r'^2}{a_{\Lambda}^2} \bigg ) \frac{\tilde{L}^2}{r'^2} .
    \end{equation}
Here we recognize the anti-gravitational potential, $V_{dS}(r') = - ({c^2} / 2) ({r'^2} / a_{\Lambda}^2)$, introduced in \Eq{comparison}. For radial time-like geodesics, having $\tilde{L} = 0$, that produces an outward acceleration, $d^2 r'/ d \tau^2 = c^2 (r'/a_{\Lambda}^2) > 0$. No massive particle can remain at rest for any $r' > 0$. Generally, it will be pushed beyond the $a_{\Lambda}$ horizon in a finite interval of `proper time,' $\Delta \tau$. This agrees with our construct of dark energy. Furthermore, an $m$-particle may stand at rest indefinitely in an unstable equilibrium position at $r' = 0$ only if it also has minimal $\tilde{E}^2 - 1 = \tilde{L}^2 = 0$. Thus, in order to keep $m_g$-photons at rest in a $g$-BEC, a non-gravitational (QM) inward pressure or `tension' must be exerted uniformly throughout their density.

For a particle with $m = 0$, I obtain
\begin{equation}\label{geodesicradialphotonaffine}
   \bigg(\frac{d r'}{d\lambda}\bigg)^2
    = \frac{E^2}{c^2}  - \bigg (1 - \frac{r'^2}{a_{\Lambda}^2} \bigg ) \frac{L^2}{r'^2} .
    \end{equation}
For radial null geodesics, having $L = 0$, $m = 0$ particles move freely, without any acceleration, $d^2 r'/ d \lambda^2 = 0$, in terms of any affine parameter, $\lambda$. In terms of `static coordinate time,' $t'$, however, we have
\begin{equation}\label{geodesicradialphoton}
   \frac{d r'}{d t'} = \pm c \bigg (1 - \frac{r'^2}{a_{\Lambda}^2} \bigg ) .
    \end{equation}
A $m = 0$ particle can then stop or stand at rest at the $a_{\Lambda}$ horizon with $L = 0$, though it takes an infinite amount of coordinate time, $t'$, to reach (or escape from) that horizon.    

All these results correspond to those of the Schwarzschild metric,\cite{Schutz2Ed, Hobson, Rindler, Resca, Rafael} on account of \Eq{comparison}. In particular, while approaching the horizon, material sources emit increasingly red-shifted light backward to the observer at $r' = 0$. That feature of \Eq{staticmetric} was soon realized and related to Slipher's 1912 discovery of galactic red-shifts.\cite{Moschella} That also provides an alternative interpretation of the gravitational red-shift in $dS_4$. Rather than viewing it as a result of the $R(t)$ expansion, $V_{dS}(r')$ generates `an inverse gravitational red-shift,' wherein incoming light in \Eq{geodesicradialphoton} must climb, in static coordinate time, $t'$, a potential hill in order to reach the observer.\cite{Ibison}  

\section{Perfect fluids}\label{Perfect fluids}

Exploiting twice-contracted Bianchi identities, Einstein constructed his $G^{\mu \nu}$ tensor in \Eq{Einstein} as divergence free, thus allowing four continuity equations that express conservation of energy and momentum for stress-energy tensor components, 

%Further contraction of those continuity equations with the fluid four-velocity, $U^{\mu}$, yields 

%
%
\begin{equation}\label{ConservationLaw}
   (T^{\mu \nu}) _ {; \nu} = 0 ,
    \end{equation}
where a semi-colon conventionally denotes covariant differentiation. 
%That \Eq{ConservationLaw} is in fact equivalent to $(T^{0 \nu}) _ {; \nu} = 0$ time-like conservation.\cite{Schutz2Ed}

In FLRW geometry, contraction of \Eq{ConservationLaw} with $U_{\mu}$ results in the elimination of $\ddot{R}$ in the two Friedmann-Lema\^{i}tre equations.\cite{Schutz2Ed, Hobson} That yields for cosmology an energy conservation equation reminiscent of the first law of thermodynamics for a perfect fluid,\cite{Schutz2Ed, Hobson, Hartle} having no viscosity nor heat conduction: 
\begin{equation}\label{TDLaw}
  \frac{d}{dt} (\rho R^3) + p \frac{d}{dt} (R^3) = 0 .
    \end{equation}

In \Eq{TDLaw} the energy density, $\rho = T_{\mu\nu} U^{\mu}U^{\nu} / c^2$, 
and pressure, $p = \frac{1}{3} T_{\mu\nu} (g^{\mu\nu} + U^{\mu}U^{\nu} / c^2)$, are defined as scalar functions under general coordinate transformations.

It is further possible to show that\cite{Schutz2Ed, Hobson}  
\begin{equation}\label{contraction}
(U^{\mu})_{;\mu}=\partial_{\mu}U^{\mu}+\Gamma_{\alpha\mu}^{\mu}U^{\alpha}=3\frac{\dot{R}}{R}
\end{equation}
in FLRW geometry. Then
\begin{equation}\label{NumberConservation}
  ( N^{\mu} ) _{; \mu} \equiv ( n U^{\mu}) _{; \mu} = \dot{n}+3 n \frac{\dot{R}}{R} ,
\end{equation}
where $N^{\mu}$ denotes the number-flux four-vector. When set equal to zero, the continuity of the four-divergence \Eq{NumberConservation} provides local conservation of particle number for a perfect fluid element in FLRW geometry.\cite{Schutz2Ed, Hobson} Einstein's field \Eq{Einstein} imply various energy conditions,\cite{Hobson, Wald, Visser} but they do not generally require conservation of particle number. 

Let us then consider a matter-energy density, $\rho_m$, with an equation of state

\begin{equation}\label{MatterDensity}
p_m << \rho_m = \frac{A}{[R(t)]^3} ,
  \end{equation}
which obeys \Eq{TDLaw} approximately.\cite{constant} That occurs when variously called `random' or `thermal' or `peculiar' relative velocities generate energies much smaller than the rest-energy mass, $m c^2 > 0$, of the particles.\cite{Schutz2Ed, Hobson, Hartle} It is exactly so for a pressure-free perfect fluid called `dust,' in which all particles are perfectly comoving with the FLRW geometry.
That suggests consideration of a particle density 
\begin{equation}\label{ParticleMatterDensity}
n_m = \frac{\rho_m}{m c^2} .
  \end{equation}
Contracting \Eq{ConservationLaw} with $U_{\mu}$, it is easy to show that \Eq{NumberConservation} vanishes and provides conservation of particle number for $n_m$ just as approximately as \Eq{TDLaw} is obeyed for $\rho_m$. 

The situation is different for radiation energy density, which obeys an equation of state\cite{constant}   
\begin{equation}\label{RadiationDensity}
 3 p_r = \rho_r = \frac{B}{[R(t)]^4} .
  \end{equation}
In that case, \Eq{TDLaw} is exactly satisfied, even though neither \Eq{PerfectFluid} nor \Eq{NumberConservation} apply, as a particle number density cannot be determined for radiation.\cite{Pathria} That is related to the fact that ordinary radiation is associated with transverse massless photons. It takes no energy to create any number of zero energy massless photons, which must then coexist with energetic photons in a two-phase equilibrium between a BEC and a `vapor' phase with $\mu = 0$ at \textit{all} temperatures. I view that as a mathematical idealization or a limiting case at best.\cite{RescaPhoton}

%Let us establish whether `fluid elements' are `comoving' or `at rest' in FLRW geometry. The freely falling comoving frame of the FLRW geometry and coordinates derives from the dynamics of the average distribution of matter in the universe. Upon the asymptotically exponential expansion of $R(t)$ in the FLRW geometry, matter density stretches and `thins' proportionally to $R^{-3}$. Galaxies recede from one another and they will eventually fade away at the EH, $a_{\Lambda}$, with perhaps some exceptions for bound local groups. By contrast, dark energy density remains constant and unaffected by the $R(t)$ expansion in the FLRW geometry, hence at rest therein. 

For the dark energy density, as given in \Eq{darkenergy}, the situation is profoundly different. In fact, $\rho_{\Lambda}$ remains a uniform \textit{universal constant} relative to the fixed $a_{\Lambda}$ horizon in both FLRW geometry and $dS_4$ static coordinates. Rather than \textit{comoving}, I thus refer to the dark energy density as \textit{at rest} in FLRW geometry. Correspondingly, a negative pressure or tension is needed to keep the dark energy \textit{density} constant in FLRW geometry, while its scale factor, $R(t)$, expands. That is completely different from the case of matter density or `dust' in \Eq{MatterDensity}, which instead \textit{comoves} and `stretches' with the $R(t)$ expansion in FLRW geometry.

These basic considerations thus provide the foundation for my model of the dark energy $g$-BEC that I develop in the next \Secref{DarkEnergy}.

\section{Dark energy as a BEC of cosmologically massive photons}\label{DarkEnergy}

For kinetically energetic photons in the ordinary vapor phase, FL equations, \Eq{firstFL} and \Eq{secondFL}, must include a radiation contribution provided by \Eq{RadiationDensity}. At the present cosmic time, $t_0$, the $R(t_0)$ expansion has already reduced that ordinary electromagnetic radiation contribution to only a small $\Omega_{r, 0}\simeq 5 \mathrm{x} 10^{-5}$ density fraction.\cite{Hobson}

Aside from that, dark energy may be modeled as follows. In a previous paper,\cite{RescaPhoton} I conjectured that cosmologically minimally massive gauge bosons such as photons - let me leave aside gluons till \Secref{gluons} - may form a BEC at rest in FLRW geometry. According to \Eq{mass} and \Eq{darkenergy}, that BEC has a number density
\begin{equation}\label{numberdensity}
n_g = \frac{\rho_{\Lambda}} {m_g c^2} = \frac{c^3 \sqrt{\Lambda} / C_g} {8 \pi h G} = \frac{\sqrt{\Lambda} / C_g} {8 \pi l_P^2} \simeq 2.6 \mathrm{x} 10^{41} m^{-3} / C_g ,
\end{equation}  
where $l_P = \sqrt{h G / c^3} = 4.051285 \mathrm{x} 10^{-35} m$ denotes the Planck length. Thus $m_g$-photons belonging to the BEC phase provide a stress-energy tensor contribution $T_{\Lambda}^{\mu \nu} = p_{\Lambda} g^{\mu \nu}$ to Einstein's \Eq{Einstein} with $T^{\mu \nu} = 0$, as prescribed in \Eq{PerfectFluidLambda} with $\Lambda >0$. 

Thus there is a fundamental difference between a perfect fluid that involves an energy density of matter or dust with conserved particle number, as derived from \Eq{MatterDensity} and \Eq{ParticleMatterDensity}, and the special perfect fluid that I associate with dark energy density, having a particle number density given in \Eq{numberdensity}. In fact, my $n_g$ does \textit{not} yield particle number conservation in \Eq{NumberConservation}, even though its equation of state in \Eq{darkenergy} exactly obeys all energy-momentum conservation laws, \Eq{ConservationLaw} and \Eq{TDLaw}. Rather, dark energy $m_g$-particles in the BEC are generated at a steady rate of
\begin{equation}\label{GenerationRate}
( N_g^{\mu} ) _{; \mu} = 3 n_g \frac{\dot{R}}{R} \simeq 3 n_g \frac{c} {a_{\Lambda}} = \frac{\sqrt{3} c} {8 \pi C_g ( l_P L_g\ )^2} ,
\end{equation}  
as $R(t)$ of the FLRW geometry expands and the Hubble parameter approaches $H_{\Lambda} = c / a_{\Lambda}$.

Equivalently, the relative rate of change of particle number is 
\begin{equation}\label{GenerationRateRelative}
\frac {( N_g^{\mu} ) _{; \mu}} {n_g} = 3 \frac{\dot{R}}{R} \simeq \frac {\sqrt{3}} {T_g} ,
\end{equation}  
where $T_g = L_g / c \simeq 10^{10}$ years. On that cosmic time scale, each $m_g$-particle must then approximately double in order to keep my $n_g$ particle \textit{density} constant overall. 

%This result clearly differs from its analogue of the continuity equation in flat Minkowski space-time for the following reason. In FLRW geometry, although fluid elements of dark energy are not accelerated along comoving geodesics, space-time itself `stretches' along other world-lines. That alone generates a non-zero current density flux, corresponding to the $R(t)$ expansion through the $a_{\Lambda}$ event horizon. While that `thins' ordinary matter and radiation densities, \Eq{MatterDensity} and \Eq{RadiationDensity}, dark energy $\rho_{\Lambda}$ and $n_g$ \textit{densities} remain constant overall. However, an $m_g c^2$ particle energy doubling must occur for each $R(t)$ doubling approximately at $T_g$ intervals. 

In order to appreciate these results more intuitively, define a space-volume $V = R^3$, which also expands at a rate $\dot{V} = 3 V \dot{R} / R$. Now the number-flux four-vector per unit space-volume, $F_g^{\mu} = N_g^{\mu} / V$, is conserved, i.e., $(F_g^{\mu});\mu = 0$. This means that $m_g$-particles must be created at exactly the same rate as space-time itself is created in FLRW geometry, in order to keep universally constant the $n_g$ particle \textit{density}, exactly as we expect from dark energy. 

Now compute that
\begin{equation}\label{ConservationLawCheck}
 U_{\mu}  (T_{\Lambda}^{\mu \nu}) _ {; \nu} = - (\rho_{\Lambda} + p_{\Lambda}) (U^{\nu}) _ {; \nu} .
    \end{equation}
This means that we have conservation of dark energy and \Eq{TDLaw} only if the equation of state $p_{\Lambda} = - \rho_{\Lambda}$ is precisely satisfied. Namely, creation of $m_g$-particles at the rate of \Eq{GenerationRateRelative} does not result in increase of dark energy only if a negative pressure balances exactly each $m_g c^2$ creation of particle energy.

How could that be fully justified in my QM-GR model? Perhaps non-linearities in a QM attractive photon-photon interaction field in the BEC extract each $m_g c^2$ quantum precisely from a corresponding space-time volume expansion. Although that can hardly be `proved' at present, I will at least refer to some prospects in \Secref{GPP}. 

\section{Flat-space estimate of photon mass and BEC inter-particle distance}\label{Flat-space}

In this \Secref{Flat-space}, I shall assume a \textit{Euclidean spatial geometry}, which applies to FLRW geometry in a flat universe with $\kappa = 0$ in \Eq{FLRWmetric}. These assumptions thus amount to the so called \textit{de Sitter universe}, which satisfies the `perfect cosmological principle,' assuming isotropy and homogeneity equally throughout space and time.\cite{wiki} Correspondingly, both FL \Eq{firstFL} and \Eq{secondFL} coincide for $T^{\mu \nu} = 0$ and $\kappa = 0$. 

Now, this spatial flatness assumption may not be entirely consistent with the $dS_4$ static metric of \Eq{staticmetric}, which has closed spatial sections.\cite{Moschella} Asymptotically, however, if not altogether, the $\Omega_{\kappa} (t)$ density parameter hardly matters. Thus, at least approximately, I will refer alternatively to desirable `near-flatness' properties of both the expanding flat slicing (Lema\^{i}tre) and the static slicing (de Sitter) of $dS_4$ that I later consider in \Secref{Zero-point}. 

Here let me then most simply assume that each particle, being almost free and virtually at rest in the BEC, is subject to a QM effective non-relativistic attractive spherical-well potential,  

\begin{equation}\label{potential}
V(r) =  \left\{
        \begin{array}{ll}
            0, & \quad  r < a_{\Lambda}, \\
            V_{0c} > 0, & \quad r > a_{\Lambda},
        \end{array}
    \right.
\end{equation}
extending the \textit{Euclidean geometry} to all space at any cosmic time.

In \Eq{potential} I then assume that $V_{0c}$ provides the minimal QM critical depth that binds just one particle state
%with $l = 0$ 
at the top of the cosmologically constant Euclidean well. By solving the time-independent Schr\"{o}dinger equation with $m{_g}$ given in \Eq{mass}, I immediately obtain that  

\begin{equation}\label{CriticalPotential}
V_{0c} = \frac{\hbar^2}{2 m_g} \bigg(\frac{\pi}{2 a_{\Lambda}} \bigg)^2 = \frac{1}{96 C_g^2}  m_g c^2 .
\end{equation}

Since there are no `walls' at $a_{\Lambda}$, that horizon must raise steeply a QM square-well potential barrier therein, generating a corresponding gradient, hence, an \textit{inward} force times a $dr$-displacement on each particle at the horizon, amounting to $V_{0c}$. I may then combine $n_g$ in \Eq{numberdensity} with \Eq{CriticalPotential} to conclude, by definition of pressure, that

\begin{equation}\label{rescaequation}
- p_{\Lambda} = n_g V_{0c} = \frac{1}{96 C_g^2}  \rho_{\Lambda} \rightarrow \rho_{\Lambda} .
\end{equation}

In the last step of \Eq{rescaequation} I recalled that the invariant continuity \Eq{ConservationLaw}, which allowed the introduction of the cosmological term in Einstein's field \Eq{Einstein} in the first place, requires that \Eq{darkenergy} holds for a perfect fluid of dark energy. Thus I set
\begin{equation}\label{rescaconstant}
C_g = 1/ \sqrt{96} \simeq 0.1 .
\end{equation}

Albeit qualitatively, this model illustrates how it is possible to constrain conjectures like those of \Eq{fundamental}, \Eq{mass} and \Eq{numberdensity} with an independent requirement, such as that of \Eq{darkenergy}. In this case, photons acquire a rest-energy mass   
\begin{equation}\label{massvalue}
m_g c^2 =  C_g h c \sqrt{\Lambda} \simeq 1.3 \mathrm{x} 10^{-33} \mathrm{eV}.
\end{equation}  

Following \Eq{numberdensity}, I thus estimate that the average distance between photons in their BEC is
\begin{equation}\label{averagedistance}
d_g \simeq  {n_g^{-1/3}}  = 2 (\pi C_g)^{1/3} \bigg[ \frac{l_P^2}{\sqrt{\Lambda}} \bigg]^{1/3} \simeq 7.5 fm .
\end{equation}  
%
%

%I may also estimate the gravitational pair-potential attraction between photons in the BEC as $V_{12} \simeq G m_g^2 /(2 d_g)$. Its ratio to $V_{0c}$ equals $(64 \pi)^{-1/3} (l_P \sqrt{C_g \Lambda})^{4/3} \simeq 10^{-82}$. This suggests that the BEC of photons is basically held together by an effectively attractive bosonic inter-particle QM correlation, while any further gravitational attraction between photons in the BEC may be negligible.
%The $V_{12}$ scale is also utterly negligible compared to that of electro-weak interactions.

Because of dimensional constraints, it may appear that \Eq{rescaequation} contains some degree of circularity. Conceptually, however, that is not the case. The first step in \Eq{CriticalPotential} derives from the fundamentally \textit{non-relativistic} version of the uncertainty principle of Heisenberg \textit{et al.} Initially, $V_{0c}$ does not contain explicitly any $c$ factor, while $m_g$ stands in the \textit{denominator}. On the other hand, $m_g$ derives from a fundamentally independent \textit{relativistic} version of the uncertainty principle in \Eq{fundamental}, containing a $c$ factor. As a result of that, a factor of $c^2$ appears in the second step of \Eq{CriticalPotential}, while $m_g$ switches to the \textit{numerator}. Thus, ultimately, \Eq{CriticalPotential} connects vastly different non-relativistic and relativistic quantum uncertainties and bridges corresponding microscopic ($h$) and cosmological ($\Lambda$) scales. The other factor, $n_g$, that enters in \Eq{rescaequation} ultimately simplifies two $m_g c^2$ factors away, yielding a numerical proportionality with $\rho_{\Lambda}$, independently of $h$. Thus, fixing the numerical proportionality constant, i.e., $C_g$ in \Eq{rescaconstant}, according to \Eq{darkenergy}, may derive from an extremely simplified model and estimate of dark energy, but conceptually that is neither trivial nor circular.

\section{Zero-point energy of the BEC of cosmologically massive photons}\label{Zero-point}

The estimate of $C_g$ in \Eq{rescaconstant} derives from QM relations expressed in \Eq{fundamental} and \Eq{CriticalPotential} in particular. The symmetry group that underlies SM-QFT is that of flat Minkowski space-time, $M(1,3)$. However, de Sitter space, $dS_4$, is an Einstein manifold with constant invariant Ricci scalar curvature, $g_{\mu \nu} R^{\mu \nu} = 4 \Lambda > 0$, even for $\kappa = 0$. That space-time curvature may be negligible even up to galactic scales, but it clearly becomes sizeable at the event horizon scale of $dS_4$, \Eq{horizon}. That is where the treatment of \Secref{Flat-space} and the result of \Eq{CriticalPotential} in particular can hardly be held as precise.

As an alternative and independent approach, let me then consider the zero-point energy of the BEC of cosmologically massive photons in de Sitter static metric, \Eq{staticmetric}, with a corresponding coordinate-time $t'$-slicing. Referring to \Eq{comparison} and \Eq{geodesicradial}, I have already discussed the quadratic \textit{repulsive} de Sitter potential, $V_{dS}(r')$, induced by the $dS_4$ anti-gravitational curvature. I then assume that in such static metric and $t'$-slicing the BEC of cosmologically massive photons is kept at rest in equilibrium by an \textit{opposite attractive} QM harmonic oscillator (HO) potential per $m_g$-mass, 
\begin{equation}\label{zerostaticpotential}
V_{0g}(r') = - V_{dS}(r') = \frac{c^2}{2 a_{\Lambda}^2} r'^2  = \frac{1}{2} \omega_{dS}^2 r'^2 ,  
\end{equation}
where $\omega_{dS} = c / a_{\Lambda}$ is the HO angular frequency.

That yields the zero-point energy density for the BEC as
\begin{equation}\label{zero-point}
\rho'_{0g} = \bigg ( \frac{1}{2} \hbar \omega_{dS} \bigg ) n'_g = \frac{1}{4 \pi \sqrt{3} C'_g} \rho_{\Lambda} \rightarrow \rho_{\Lambda}.
\end{equation}
Here I define again $m'_g$ and $n'_g$ as in \Eq{mass} and \Eq{numberdensity}, but the numerical coefficient, $C'_g$, now differs from the previous $C_g$ that is determined in \Eq{rescaconstant}. Two $h$ factors cancel in \Eq{zero-point} and $\rho'_{0g}$ is remarkably independent of $h$. Thus, if I set
\begin{equation}\label{rescasecondconstant}
C'_g = \frac{1}{4 \pi \sqrt{3}} \simeq 0.045944, 
\end{equation}
I can identify the dark energy density, $\rho_{\Lambda}$, with the zero-point energy density of the BEC, $\rho'_{0g}$. 

Evidently, \Eq{zero-point} and \Eq{rescasecondconstant} imply that
\begin{equation}\label{zero-pointBIS}
\epsilon'_g = m'_g c^2 = \frac{1}{2} \hbar \omega_{dS} = C'_g h c \sqrt{\Lambda} \simeq 0.59 \mathrm{x} 10^{-33} \mathrm{eV} ,
\end{equation}
and conversely. This confirms that the $g'$-photon rest-energy mass corresponds to the zero-point energy of the attractive QM-HO potential that balances $V_{dS}(r')$.

From an order-of-magnitude perspective, switching between my numerical coefficients hardly matters, since $C'_g$ has only about half the value of $C_g$. By comparison, zero-point energies resulting from standard QFT calculations may overestimate vacuum fluctuations by as many as 120 orders of magnitude.\cite{Bull, Durrer, Steinhardt, Dadhich, BianchiRovelli, Weinberg, wikizero, Mandl, Ballentine, Hobson, Visser} In fact, such calculations typically involve quantum field frequencies up to an ultra-violet cut-off at the Planck scale, whereas \Eq{zero-point} requires only a single frequency, $\omega_{dS}$, associated with an infra-red cosmological cut-off at the horizon scale. I will return to discuss that in \Secref{Units}.
 
%It also follows that 
% 
%
%\begin{equation}\label{gravity-zero-point}
%\frac{G}{r'} \bigg ( \frac{\rho'_{0g}}{c^2} \frac{4 \pi}{3} {r'}^3 \bigg ) =  \frac{1}{4 \pi %\sqrt{3} C'_g}  \bigg ( \frac{1}{2} \omega_{dS}^2 r'^2 \bigg )  .
%= - \frac{1}{2} V_{0c} .
%\end{equation}  
%
%
%This means that the classical gravitational potential with zero-point energy coincides with the BEC opposite of de Sitter anti-gravitational potential, per particle-mass, for \textit{all} $r'$-distances,  up to the Schwarzchild and de Sitter horizons. Remarkably, all $h$ factors cancel in the former. Therefore, the classical $\Lambda$-CDM equation of state for the negative pressure or `tension', \Eq{pressure}, still follows from \Eq{gravity-zero-point}.

%This indicates that, although not exact, the $m_g$ estimate of the photon mass in \Eq{massvalue} may be correct within a few orders of magnitude, if not a single one. 

%Major problems also arise because QFT predictions typically do not include curvature effects of space-time nor consideration of the cosmological constant, $\Lambda$, in particular. I will return to that point when comparing Planck and cosmological units in \Secref{Units}.

\section{Constant dark energy density and gravitational collapse}\label{DarkPressure}

%Although not apparently consistent with my previous formulation of the dark energy-pressure relation, nor with itself for that matter, 
It is instructive to further consider a classical GR perspective of dark energy condensation in terms of impending gravitational collapse. 

Thus consider a spherically symmetric and uniform matter density, $\rho_{\Lambda} / c^2$, extending initially in a spatial Euclidean geometry. Consider the radial function
\begin{equation}\label{Schwarzschild}
f(R') = 2 \frac{G}{c^4} \rho_{\Lambda} \frac{4 \pi}{3}  R'^3 - R' .
\end{equation}  
Gravitational collapse is bound to occur when the radius, $R'$, of $\rho_{\Lambda}$ reaches a Schwarzschild radius, $R'_S$, such that $f(R'_S) = 0$. This type of basic argument is often used to estimate gravitational collapse of black holes, galaxies and other astronomical objects.\cite{Rindler} 

For dark energy, as given in \Eq{darkenergy}, one immediately obtains that 
\begin{equation}\label{collapse}
f(R'_S) = 0 \Longrightarrow R'_S = a_{\Lambda} . 
\end{equation}  
This is a remarkably interesting result, independent of any further theory or supposition. It is related to the fact that the $a_{\Lambda}$ horizon appears explicitly in the static metric of $dS_4$ and that static metric is formally similar to the Schwarzschild metric, as shown with \Eq{comparison}.

%In Euclidean geometry, the energy of the distribution is 
% 
%
%\begin{equation}\label{Energy}
%E = \rho_{\Lambda} \frac{4 \pi}{3}  R'^3 .
%\end{equation}  
%
%
%The radially attractive force, $- dE/dR'$, yields a negative pressure or tension, 
% 
%
%\begin{equation}\label{pressure}
%p_{\Lambda}= - \rho_{\Lambda}, 
%\end{equation}  
%constant throughout the energy density distribution. 

%At the Schwarzschild radius, the Newtonian gravitational potential per particle-mass becomes 
% 
%
%\begin{equation}\label{potentialenergy}
%W(R'_S) = - \frac{G}{R'_S} \bigg ( \frac{\rho_{\Lambda}}{c^2} \frac{4 \pi}{3} {R'_S}^3 \bigg ) = - \frac{1}{2} c^2 .
%= - \frac{1}{2} V_{0c} .
%\end{equation}  
%
%
%Curiously, such an elementary gravitational collapse treatment of dark energy condensation, combined with a classical Newtonian treatment of binding of a corresponding $m_g$ particle at the $R'_S = a_{\Lambda}$ horizon, turn out to coincide, 

%Except for numerical factors, \Eq{potentialenergy} is consistent with the non-relativistic QM formulation of $m_g$ binding, obtained in \Eq{CriticalPotential}. 

%However, \Eq{potentialenergy} holds for a test particle of any mass.  So, in and of itself, \Eq{potentialenergy} cannot specify any value for $m_g$, let alone the cosmological and QM value derived in \Eq{massvalue}. 

Qualitatively, these results may indicate that relativistic QM, \Eq{fundamental}, non-relativistic QM, \Eq{CriticalPotential}, and classical GR, \Eq{collapse}, provide at least compatible accounts of dark energy condensation within de Sitter horizon, bridging a chasm between corresponding nuclear and cosmological scales, as defined by $d_g$ and $a_{\Lambda}$. %If one considers that these perspectives bridge a chasm between nuclear and cosmological scales, one can hardly be dismissive of the idea that confinement of a real elementary particle to a cosmological event horizon sets the infrared limit of the particle's energy and mass.

%There are well-known theories of gravitational vacuum and dark energy stars.\cite{Mazur, Chapline, Laughlin} Advanced formulations of quantum phase transitions to BE condensation of dark energy at a cosmological scale on the verge of gravitational collapse may be much further developed with such expertise.

\section{Cosmic Microwave Background}\label{CMB}

Description of dark energy as a BEC of cosmologically massive photons at rest in de Sitter space involves consideration of statistical properties of equilibrium between such BEC phase and a vapor phase of $m_g$ photons. Analysis of cosmic microwave background (CMB) data may inform or constrain that description.

One should also be mindful of the so-called `strong energy condition,' requiring $\rho \ge - 3 p$ for the equation of state.\cite{Hobson, Wald, Visser} Most notably, that does not hold for dark energy in the standard $\Lambda$-CDM, while it should apply to a standard BEC in a Euclidean or Minkowskian space-time. On the other hand,
%Careful consideration of all these facts does not reveal, however, any fundamental contradiction. The $g$-BEC differs from a standard BEC at least in the following respects. 
a standard BEC is confined by an external volume, whereas the condensate that I envision, to which I refer more specifically as a $g$-BEC, must be self-confining within the de Sitter horizon. 

Furthermore, $m_g$-photons in the ordinary vapor phase do not need to condense directly into the $g$-BEC. Rather, they may maintain that equilibrium via interactions with other particles, such as absorption, emission or scattering by electrons and protons, or even neutrinos. In fact, it is generally possible to investigate electro-weak processes including a massive photon propagator.\cite{Mandl, Liang, Ruegg, Heeck, Reece, Nieto} Such calculations are quite involved, however, even to low perturbative orders. 

Although I may not enter into any such technical issues at this time, let me at least note yet another `curious coincidence.' The low-energy elastic scattering of photons by free electrons results in Thomson cross section, proportional to the square of the classical radius of the electron, $r_0 = e^2 / m_e c^2 \simeq 2.818 fm$. In value, that is remarkably close to the average distance between photons in the $g$-BEC, which is given in \Eq{averagedistance} as $d_g \simeq 7.5 fm$. For all that is currently known (but see Postscript, \Secref{Postscript}) the electron charge, $e$, that enters $r_0$ seems unrelated to the fundamental constants $h, c, G, \Lambda$ that enter $d_g$, and yet $r_0 \sim d_g$. Somehow the $fm$-scale seems to be the microscopic length of convergence not only of classical and quantum electromagnetic and nuclear interactions, but also of $g$-BEC interactions derived from a cosmological horizon.\cite{RescaPhoton}

In any case, relative corrections derived from a photon mass, $m_g$, to QFT scattering cross sections in non-relativistic limits are at most of the order of $m_g / m_e \sim 3 \mathrm{x} 10^{-39}$. Such relative corrections to atomic spectral lines up to the UV/X-ray range become at most of the order of $\alpha^{-2} m_g / m_e \sim 5 \mathrm{x} 10^{-35}$. Any such corrections appear too small to be directly observable at present.

Let me then address at least some questions of compatibility between $g$-BEC equilibrium and CMB observations for `ordinary' photons at a most basic level of approximation.

The $g$-BEC QM-average of momentum vector, \textbf{p}, of $m_g$-photons is $< \textbf{p} >_g = 0$. %although its variance is $< p^2 >_g = 2 (m_g c)^2$.
Consistently with \Eq{potential} and \Eq{CriticalPotential}, we may thus assume for the vapor phase a dispersion relation  
\begin{equation}\label{energy-momentum}
\epsilon = \sqrt{(m_g c^2)^2 + c^2 p^2} - m_g c^2 ,
\end{equation}
between kinetic energy, $\epsilon$, and average momentum magnitude, $p = <|\textbf{p}|>$. Thus $\epsilon$ excludes a zero-point energy, $V_{0c} = m_g c^2$, and varies in the range $(0, +\infty)$.

After bringing $m_g c^2$ to the left-hand side of \Eq{energy-momentum} and squaring the result, the ratio between $d p$ and $d \epsilon$ can be obtained in terms of $\epsilon$. From that the exact expression for the density of states,
\begin{align}\label{DOS}
g(\epsilon) d \epsilon  = g_S (V/h^3) 4 \pi p^2 ( d p / d \epsilon) d \epsilon
  = g_S \frac{V}{(h c)^3} 4 \pi \sqrt{\epsilon} \sqrt{\epsilon + 2 m_g c^2} (\epsilon +  m_g c^2) d \epsilon ,
\end{align}
can be easily derived. The spin degeneracy for $m_g$-photons must be $g_S = 3$, while it is $g_S = 2$ for massless transverse photons.

The average kinetic energy of $m_g$-photons in the vapor phase can thus be obtained as 
\begin{equation}\label{energy}
U = < E > = \int_{0}^{\infty} \frac{\epsilon g(\epsilon) d \epsilon}{e^{\beta \epsilon} - 1}.
\end{equation} 

Approximate integration in a small initial $(0, m_g c^2)$ interval contributes negligibly to \Eq{energy}. Thus an expansion for $m_g c^2 << \epsilon$ can be assumed to hold for the entire integral. The resulting correction to $U_P$ as provided by Planck's law for massless photons then is
\begin{equation}\label{energycorrection}
\frac{\Delta U}{U_P} \simeq \bigg( \frac{3}{2} \bigg) \bigg( \frac{2 \zeta (3)}{3 \zeta(4)} \bigg) \bigg( \frac{m_g c^2}{k_B T} \bigg),
\end{equation} 
where $\frac{2 \zeta (3)}{3 \zeta(4)} \simeq 0.74$. Independently of that, the first $(3/2)$ factor derives from the assumption of $g_S = 3$ for massive ($g_S = 2$ for massless) photons in the numerator (denominator) of \Eq{energycorrection}.

%Notice that the ratio in \Eq{energycorrection} is independent of whatever value may be assumed for $g_S$.

The current temperature of the CMB is $T_0 \simeq 2.725 K$, corresponding to $k_B T_0 \simeq 2.35 \mathrm{x} 10^{-4}eV$. Comparison of that $k_B T_0$ value with the $m_g c^2$ value obtained in \Eq{massvalue} yields a relative correction to Planck's law for the current CMB of $\Delta U / U_P (T_0) \simeq 6.2 \mathrm{x} 10^{-30}$, which seems currently undetectable. At the recombination epoch of the optical horizon, $T_{rec} \simeq 3000 K$ further reduced $\Delta U / U_P$ by another factor of $T_0 / T_{rec} \simeq 9 \mathrm{x} 10^{-4}$, thus yielding $\Delta U / U_P (T_{rec}) \simeq 5.6 \mathrm{x} 10^{-33}$. So, there may not be any directly detectable deviation from current measurements of the CMB caused by any such tiny attribution of $m_g c^2 \simeq 1.3 \mathrm{x} 10^{-33} \mathrm{eV}$ to the actual photon rest energy mass. However, the $n_g$ density of $m_g$-photons in the $g$-BEC is extremely high, as given in \Eq{numberdensity}. Thus, measurable effects of the $g$-BEC on other kinds of observations are entirely possible, if not likely, as I will mention in the next \Secref{GPP}. 

In any case, a spin degeneracy factor of $g_S = 2$ must be attributed to transverse massless photons for Planck's law to describe the CMB with such a remarkable precision. However, massive photons must have $g_S = 3$, no matter how light they may be cosmologically. This apparent discrepancy must then be associated with a mathematical singularity in dealing with a $m_g \rightarrow 0$ `sick' limit vs. setting $m_g = 0$ in the first place. This problem in fact recurs in QFT.\cite{Mandl, Liang, Ruegg, Heeck, Reece, Nieto} For the moment, I propose to regard the vapor phase as essentially or overwhelmingly composed of transverse massless photons, having $g_S = 2$. By contrast, the $g$-BEC phase, having no wavevector direction, since $< \textbf{p} >_g = 0$, 
%or $p=<|\bar{p}|>_g = 0$, 
makes no distinction between transverse and longitudinal photons, corresponding to a `vacuum ground state' to which $g_S = 3$ applies.

Lately I became aware that BE condensation of photons in an optical microcavity has been achieved.\cite{Klaers} That experimental realization is formally equivalent to an ideal gas of massive bosons having an effective mass of about $4 \mathrm{eV}/ c^2$, far greater than my estimate in \Eq{massvalue}. Though fundamentally important and suggestive, how such experimental findings may or may not be related to my considerations remains to be established.

\section{Gross-Pitaevskii-Poisson systems and relativistic generalizations}\label{GPP}

At this point I must at least refer to great advances that have been made in the study and terrestrial realizations of BE condensates, such as those of superfluids and trapped-cooled dilute gases.\cite{Stringari, Pethick} A vast literature has further developed regarding astrophysical and cosmological applications. Excellent papers and reviews provide relatively concise and accessible entry points to that field.\cite{Calmet, ChavanisMass, Chavanis, Hwang, Hui, Morikawa, Nishiyama, Ferreira, LeeA} Some include most lucid derivations and complete accounts of the development of the field, starting from the original hydrodynamical formulations of quantum mechanical wave equations.\cite{ChavanisMass, Chavanis} They then focus on Gross-Pitaevskii-Poisson (GPP) systems and their relativistic generalizations. Their emphasis is on dark matter related to BEC of ultra-light particles with masses of the order of $m \sim 10^{-24} eV/c^2$, although a cosmological limit of $m \sim 10^{-33} eV/c^2$ may ultimately be attained.

From my perspective, a starting point may be provided by Eqs. [27-29] of Ref. \onlinecite{Hwang}. Although those Eqs. [27-29] apply to a pressureless gas in the non-relativistic limit, they can account for the cosmological expansion of $dS_4$ in particular. Exclusive consideration of a constant $\rho_{\Lambda}$ eliminates from Eq. [28] the effects of both the quantum potential arising from the Heisenberg uncertainty principle and the quantum scattering pressure that is typical of GPP systems with a corresponding scattering length. In fact, both such quantum effects are assumed to become negligible at the cosmological scale.\cite{ChavanisMass} A term involving the gradient of a constant pressure, such as $p_{\Lambda}$ that I assume in my \Eq{darkenergy}, would also vanish in the Euler-like Eq. [28] of Ref.~[\onlinecite{Hwang}]. Based on such simplifying assumptions, my solution to the hydrodynamic Eq. [28] and its corresponding continuity Eq. [27] provides a gravitational potential that is half that of the $dS_4$ cosmologically expanding Poisson Eq. [29] of Ref. \onlinecite{Hwang}. This shows again that at least the orders of magnitude in my $dS_4$ $g$-BEC theory of dark energy are consistently viable.

A more accurate relativistic generalization of GPP systems should take into account at least some quantum potential and scattering pressure non-linear effects.\cite{Chavanis, Hwang, Hui, Morikawa, Nishiyama} In my $g$-BEC that may take perturbatively advantage of a short scattering length, since scattering of light by light begins only at fourth-order for Feynman diagrams in standard QFT.\cite{Mandl, Liang} Thus it may be possible to determine whether the fundamental equation of state of the standard $\Lambda$-CDM model, \Eq{darkenergy}, can be satisfied for a $g$-BEC at least in Minkowskian space-time. On the other hand, full extension of $g$-BEC to the cosmological scale may eventually require full consideration of QFT in curved space-times.\cite{WaldCurved, WaldHolland, Ford, Parker, BirrellCurved, Aldrovandi} Such applications may involve profound conceptual and technical complexities, even for a `simple' Einstein manifold such as $dS_4$. 

Whatever the case, it is worth noting that correct evaluation of the low-energy cross section of photon-photon scattering yields an estimate of mean-free-path for visible light in the CMB frame of about $7 \mathrm{x} 10^{52} \mathrm{ly}$, at least $10^{42}$ times greater than the size of the `observable' universe.\cite{Liang} I estimated in Eqs. [6-8] of Ref. \onlinecite{RescaPhoton} the critical temperature and number density of the BEC of $g$-photons, implying that the latter is about $6 \mathrm{x} 10^{32}$ times greater than the current CMB photon density. Thus my $g$-BEC should still remain largely transparent to visible light, by at least a $10^{9}$ scale factor. On the other hand, my $g$-photons must also gravitate and clump as dark matter around ordinary matter. Thus it may be possible to observe some dimming of most distant objects when their radiation grazes or lenses around compact objects in particular. Precise models and estimates of such effects lie beyond the scope of this paper, evidently, but at least in principle they are quite calculable, since most parameters relating to $g$-photons and the $g$-BEC medium are fairly narrowly specified.\cite{RescaPhoton}  

\section{Quantum-mechanical and cosmological time}\label{Time}

As much as anything else, conceptions of time differ profoundly in GR and QM. Precise definitions of the relativity of time are hallmarks of GR, whereas a conception of `universal time' still underlies QM.\cite{Chapline, Laughlin} On the other hand, `timelessness' appears in the Wheeler-DeWitt equation, perhaps a first step toward a quantum theory of gravity.\cite{Weinberg, MTW, Wheeler, Rovelli, RovelliReal}

Let us return, however, to Weyl's postulate and the time-like world-lines of `fundamental observers' comoving with the FLRW geometry.\cite{Hobson} Ideal `atomic' clocks carried by fundamental observers can thus be synchronized and all measure the same `proper' or `synchronous' or `cosmic' time. Thus, at least within the idealization of the cosmological principle, one may identify the cosmic time of the entire universe with the `universal time' that may apply to QM at a corresponding cosmological scale. That is indeed what \Eq{fundamental} assumes, implying simultaneous delocalization and entanglement of photons uniformly within the cosmological horizon.

It is still necessary, however, to specify the causal sequence of events involved in the picture that I propose. My basic assumption is that the photon mass, $m_g$, is created at the big bang. That creates the cosmological constant, $\Lambda$, by seeding \Eq{fundamental} and \Eq{mass} at the big bang, regardless of the fact that de Sitter space, $dS_4$, with its ultimate event horizon, $a_{\Lambda}$, is bound to dominate the universe evolution only at a much later epoch. In other words, it is $m_g$ that `originates' $a_{\Lambda}$ in \Eq{mass} and \Eq{horizon}, rather than the other way around, as \Eq{fundamental} may formally suggest, based on the uncertainty principle. Likewise, the $g$-BEC of photons also forms at a very early time of the order of $10^{-6}$ s, as estimated in Ref. \onlinecite{RescaPhoton}, regardless of the fact that the CMB develops at a much later optical horizon, around 370 kyr after the big bang. Similarly, the non-relativistic QM localization reflected in \Eq{CriticalPotential} is also seeded by $m_g$ at the big bang, regardless of the fact that the universe expansion is originally singular. It is only for descriptive purposes that I may envision the generation of $m_g$ as arising from QM confinement within $a_{\Lambda}$, simultaneously and independently of time. 

\section{Cosmological versus Planck units and uncertainty relations}\label{Units}

Now let me introduce an alternative system of cosmological units, or `$g$-units,' that complements the fundamental system of Planck units in various ways, including uncertainty relations to which either set of units inherently correspond.

Planck units fundamentally include Planck length, $l_P \equiv (\Delta l)_P = \sqrt{h G / c^3} = 4.051285 \mathrm{x} 10^{-35} m$, Planck mass, $m_P = h /(c l_P) = 3.060368 \mathrm{x} 10^{28} e V / c^2$, and Planck time, $t_P \equiv (\Delta t)_P = l_P / c = 1.351385 \mathrm{x} 10^{-43} s$.

Formally, one may combine these fundamental constants to form `relativistic uncertainty relations'
\begin{equation}\label{uncertaintyPlanck}
(\Delta l)_P ( m_P c ) = (\Delta t)_P (m_P c^2) = h, 
\end{equation} 
stemming from a Compton wavelength interpretation of the exact relation between $l_P$ and $m_P$.

Conversely, I define $g$-units of length, $L_g \equiv (\Delta L)_g \equiv 1 /\sqrt{\Lambda} \simeq 10^{10} \mathrm{ly} \simeq 0.946 \mathrm{x} 10^{26} m$, mass, $M_g \equiv h /(c L_g) \simeq 1.3 \mathrm{x} 10^{-32} e V / c^2$, and time, $T_g \equiv (\Delta T)_g \equiv L_g / c  = 1 / \sqrt{3} \omega_{dS} \simeq 10^{10} \mathrm{years} \simeq 3.2 \mathrm{x} 10^{17} s$.
%where $\mathrm{ly} \simeq 0.946 \mathrm{x} 10^{16} m$ .   

I may then combine these other fundamental constants to form relativistic uncertainty relations
\begin{equation}\label{g-uncertainty}
(\Delta L)_g ( M_g c ) = (\Delta T)_g (M_g c^2) = h .
\end{equation} 

Planck and $g$-units share the fundamental constants $c$ of special relativity (SR) and $h$ of QM. Planck units add to $c$ and $h$ the universal gravitational constant, $G$. Instead of $G$, $g$-units add to $c$ and $h$ the cosmological constant, $\Lambda$. Both $\Lambda$ and $G$ rightfully belong to Einstein's field \Eq{Einstein}. While $\Lambda$ purely defines an inverse square length for any coordinate system, $G$ requires the introduction of $h$ to generate a fundamental length, i.e., $l_P$. Thus, all three Planck units require the presence of $h$, whereas only $M_g$ requires that $h$ in $g$-units, in order to form a Compton wavelength as $L_g$. Thus, Planck units characterize the Planck era at high energies of the universe at its microscopic beginning, whereas $g$-units characterize the de Sitter era at low energies as the universe demises cosmologically. 

Thus both $G$ and $\Lambda$ are very small on corresponding scales. That suggests taking their product to form a minimal fundamental a-dimensional ratio,
\begin{equation}\label{fundamentalratio}
\sqrt \frac{h G}{c^3}  \sqrt \Lambda = \frac{l_P}{L_g} =  \frac{t_P}{T_g} = \frac{M_g}{m_P} \simeq 4.2 \mathrm{x} 10^{-61} .
\end{equation}  

There are some interesting alternative interpretations of \Eq{fundamentalratio}. For example, the inverse ratio
\begin{equation}\label{inversefundamentalratio}
\frac{L_g}{l_P} = 1 + z_{\Lambda} \simeq 2.4 \mathrm{x} 10^{60} 
\end{equation}
may be interpreted as the maximal red-shift from the $l_P$ to the $L_g$ scale, at the `look-back time' from $t_P$ to $T_g$.

Let me further consider the Casimir electromagnetic vacuum energy density with a cut-off at the Planck length, which is readily evaluated as $\rho_{CEM} = \pi h c / l_P^4$, following the derivation of Eq. [19.37] on p. 534 of Ref. \onlinecite{Ballentine}, for example. The ratio of $\rho_{CEM}$ to my $\rho'_{0g} \rightarrow \rho_{\Lambda}$ in \Eq{zero-point} is 
\begin{equation}\label{Casimir-ratio}
\rho_{CEM} / \rho_{\Lambda} =   8 \pi^2 (L_g/l_P)^2 \simeq 4.55 \mathrm{x} 10^{122} .
\end{equation}  
This ratio illustrates the discrepancy between vacuum energy densities according to standard QFT calculations\cite{Bull, Durrer, Steinhardt, Dadhich, BianchiRovelli, Weinberg, wikizero, Ballentine, Mandl, Hobson, Visser} and my estimate of $\rho'_{0g}$ with a cosmological cut-off at $L_g$, which is as large as anyone may get. 

Correspondingly, let me consider 
\begin{equation}\label{Archimedes}
n'_g \bigg ( \frac{4 \pi}{3} a_{\Lambda}^3 \bigg ) = \frac{\sqrt{3}}{2 C'_g} \bigg ( \frac{L_g}{l_P} \bigg )^2 \simeq 1.1 \mathrm{x} 10^{122} ,
\end{equation}  
where $C'_g$ is given in \Eq{rescasecondconstant}. This may be interpreted as the total number of $g$-photons in the BEC that fill and still keep together the `observable' universe up to its ultimate event horizon.%\cite{Postscript}

Thus, switching between relativistic uncertainty relations at the most microscopic and cosmological scales, as expressed in \Eq{uncertaintyPlanck} and \Eq{g-uncertainty}, respectively, may provide complementary views of the QM-GR `beginning' and `demise' of the universe, spanning all of the `observable' space-time. Through it all, this causally connected universe has always been and will always remain an indivisible `atom,' as originally conceived by Lema\^{i}tre and always prescribed by QM and GR combined. 
%as it may be envisioned.\cite{Chapline, Laughlin, MTW, Wheeler} 

Planck and $g$-uncertainty relations may also imply that the heaviest `elementary particle,' born at the beginning of the universe, has a mass of the order of $m_P$, while the lightest elementary particle, surviving at its end, has a mass of the order of $M_g$. That `ultimate survivor' is the photon, I propose, which is just about as light, common and easy to produce as ultimately allowed.

%On the other hand, \Eq{uncertaintyPlanck} may suggest that at the Planck era a proto-particle with a mass of the order of $m_P$ may have formed some quantum gravity analogue of a de Sitter space with a horizon of the order of $l_P$, yielding an accelerated expansion perhaps consistent with inflation theories.\cite{Hobson} Spontaneous decay of that particle in a time of the order of $t_P$ may have ended that inflation and Planck era. Although this is mere speculation, it could be tested in terms of inflation scenarios at least qualitatively. 

\section{Cosmological mass of neutrinos}\label{Neutrino}

Standing between so vastly different $M_g$ and $m_P$ extreme values, one may wish to find some other characteristic mass. Thus, rather than taking the product, one may wish to consider the ratio between those two fundamental constants of GR, $\Lambda$ and $G$, and then form another QM mass, namely, the geometric mean
\begin{equation}\label{geometric}
\sqrt {M_g m_P} = \bigg [ \frac{h^3 \Lambda}{c G} \bigg ]^{1/4} \simeq 2 \mathrm{x} 10^{-2} e V / c^2 \simeq m_{\nu} .
\end{equation}  

Thus, in units of $e V / c^2$, we have 
\begin{equation}\label{massrange}
M_g \simeq 1.3 \mathrm{x} 10^{-32} < m_{\nu} \simeq 2 \mathrm{x} 10^{-2} <  m_P \simeq 3.06 \mathrm{x} 10^{28} .
\end{equation}  

While $M_g$ and $m_P$ masses span a range of about $60$ orders of magnitudes, their $\sqrt {M_g m_P}$ geometric mean falls close to the logarithmic middle of that range, curiously matching current estimates of neutrino masses almost within a single order of magnitude! Chances that all this happens again entirely by coincidence seem remote. Beside $m_g$-photons, if those exist, neutrinos are the lightest known fermions. They are electrically neutral and only weakly interacting. It is thus at least plausible that neutrino masses are most directly related to both GR and QM theories, perhaps fundamentally deriving from \Eq{geometric}. The central position of neutrinos in both GR and elementary particle physics has been long foreseen.\cite{Wheeler}   

Unlike \Eq{fundamental}, which is grounded on a fundamental uncertainty principle of QM, extended to further involve only $\Lambda$ in GR, \Eq{geometric} is not as simply related to that. However, the ratio of $\Lambda$ and $G$ \textit{is} fundamentally related to $\rho_{\Lambda}$ in \Eq{darkenergy}. Namely,   
\begin{equation}\label{neutrino}
c^2 \bigg [ \frac{h^3 \Lambda}{c G} \bigg ]^{1/4} = \bigg [ 8 \pi (h c)^3 \rho_{\Lambda} \bigg ]^{1/4} \simeq 2 \mathrm{x} 10^{-2} e V ,
\end{equation}  
independently of any supposition. Though not all that shines is gold, \Eq{geometric} and \Eq{neutrino} look at least as beautiful to my eyes. They may hint at a \textit{weakly-interacting $g$-photon-neutrino plasma} that may hold the universe together or provide a component of dark matter, perhaps.

%Weakly interacting $m_g$-photons and neutrinos\cite{Ruegg} may form a plasma, and \Eq{geometric} may relate to that. Currently this is mere speculation, but specific theories could be developed. 

Also consider that 
\begin{equation}\label{geometric-time}
\sqrt {t_P T_g} = \frac{h}{c^2 \sqrt{M_g m_P}} =  \bigg [ \frac{h G}{c^7 \Lambda} \bigg ]^{1/4} \simeq 2.1 \mathrm{x} 10^{-13} s
\end{equation}  
turns out to be closely related to the time of the electro-weak phase transition at about $10^{-12} s$ in the chronology of the universe.\cite{wikichronology} So, the cosmological constant, $\Lambda$, originally introduced to account for the largest scale of cosmic geometrodynamics in GR, may also be involved, upon taking the geometric mean of the corresponding $T_g$ with the Planck time, $t_P$, with the quintessential Higgs mechanism of electro-weak gauge symmetry breaking in QFT.\cite{Bertone, Mandl}  

Introduction of the $M_g$ unit further suggests to reconsider of a famous dictum not to ask `why is gravity so feeble,' but rather `why is the \textit{proton}'s mass so small?' That is, compared to Planck mass.\cite{Wilczek, WilczekTwo, wikiWilczek} However, compared to $M_g$, the proton's mass, $m_p \simeq 938 \mathrm{MeV} / c^2$, is very large! In fact, while $m_p$ still lies well within the vast range of $M_g$ and $m_P$ values, it already stands about $11$ orders of magnitudes above the value of $\sqrt {M_g m_P}$ in \Eq{geometric}, undoubtedly because of other quantum chromodynamics (QCD) elements and interactions. Thus, `scaling mount Planck' may have to start at a `zero-point base camp,' as from \Eq{zero-point} at the `bottom!'

In fact, let us consider the next weighted geometric mean
\begin{equation}\label{proton}
[ M_g m_{P}^2 ]^{1/3} \simeq 230 M e V / c^2 \simeq m_p/4.1 .
\end{equation}  
This is precisely what sets the inter-particle distance in the $g$-BEC almost at the fermi scale in \Eq{averagedistance}, i.e., $d_g \simeq 5.66 \lambda_p \simeq 7.5 fm$, where $\lambda_p \simeq 1.32 fm$ is the Compton wavelength of the \textit{proton}. Namely, via \Eq{fundamental}, the fundamental constants of GR and QM combine in \Eq{numberdensity} to yield the volume $L_g l_P^2$, setting apparently unrelated $d_g$ and $\lambda_p$ lengths at about the same $fm$-scale. However `coincidental,' this is perhaps the most tantalizing or consequential `occurrence' that I remarked both in this and in my previous paper.\cite{RescaPhoton} In fact, the dark energy density in \Eq{darkenergy} can be expressed fundamentally as
\begin{equation}\label{darkenergyfundamental}
\rho_{\Lambda} = \frac{hc} {8 \pi L_g^2 l_P^2} \simeq 3.38 \mathrm{x} 10^{9} eV m^{-3} .
\end{equation}  

\section{Massive gluons and gravitons}\label{gluons}

Gauge invariance does \textit{not} preclude the photon from acquiring mass. Beside Higgs mechanisms, a St\"{u}ckelberg mechanism applied to an Abelian hypercharge group, $U(1)_Y$, can provide photons (or other $U(1)_{B-L}$ gauge bosons) with mass.\cite{Ruegg, Heeck, Reece} In fact, it should be harder to `prove' that the photon is exactly massless than to admit that it may have a small mass, since only the latter can afford \textit{technical naturalness}.\cite{Heeck, Reece} The problem for the latter is rather to determine what value that `small mass' may actually have!

To generalize St\"{u}ckelberg mechanisms to non-Abelian gauge theories, while still preserving their unitarity and renormalizability, is far more complicated. However, there are proposals to do that and avoid the plague of almost intractable infrared divergencies in QCD, for example.\cite{Ruegg} 

Furthermore, even if QCD requires massless gluons, that applies only to flat Minkowski space-time. My proposal of \Eq{fundamental} refers instead to a cosmologically curved space-time. I maintain that \Eq{fundamental} is consistent with fundamental principles of both QM and GR applied at a cosmological scale. Those are the uncertainty principle, fundamental to QM, and a finite maximum/supremum speed, $c$, hence causality-limited horizons, fundamental to SR and GR. Thus, a cosmologically minimal mass, such as $M_g$, should limit theoretically massless gluons in the SM as well.\cite{RescaPhoton}

Now, in QED, photons are neutral, whereas in QCD gluons carry color and strongly interact with themselves. Thus, the far greater complexity of QCD interactions and vacuum fluctuations render inapplicable to gluons any elementary argument that I have put forward in this paper to further estimate $m_g$ values for the mass of the photon. Thus, gluon masses may well be higher than predicted by the cosmological $M_g$ minimum.

On the other hand, it remains tempting to maintain some arguments of Ref. \onlinecite{RescaPhoton} regarding the nearly asymptotic freedom `coincidence' already noted after Eq. [5] therein, as well as the the critical temperature of BEC in Eqs. [6-7], reaching to the end of the quark epoch, at about $10^{-6}$s and $10^{12}$K. Far more advanced theories have also considered that $\Lambda$ may be fundamentally related to masses of elementary particles such as pions or Higgs bosons,\cite{Prat} or that cold dark matter may derive from a gluonic BEC in anti-de Sitter space-time.\cite{GCT}

Remarkably, recent measurements of the positive muon anomalous magnetic moment suggest new physics beyond the SM.\cite{muon} From my perspective, one could at least consider QCD calculations on a lattice that may determine parametric values of gluon masses starting from the minimal $M_g$ that I estimated. 

Furthermore, some of the elementary argument that I have put forward in this paper may also apply to $g$-gravitons.\cite{RescaPhoton} Starting with the Pauli-Fierz approach (1939), theories of graviton mass have a long history, although that is still complicated and far from resolution.\cite{Nieto, wikiGraviton}

Famously, Einstein repeatedly failed to correctly predict any property of elementary particles exclusively from GR theory, although his `mistakes' notoriously led to most fruitful developments, such as that of the Einstein-Rosen `bridge,' for example.\cite{Rafael, Wormhole, EinsteinRosen, haghani2022compact} In the starkest of contrasts, standard QFT has proved immensely successful in predicting all kinds of properties of elementary particles, including many mass relations, without any regard to GR. If or when QFT, and mechanisms of symmetry breaking in particular, may further account for space-time curvature of GR,\cite{WaldCurved, WaldHolland, Ford, Parker, BirrellCurved, Aldrovandi} then the universal and cosmological constants $G$ and $\Lambda$ may be finally proved to enter fundamental relations such as those of photon and neutrino masses that I have barely estimated.

%Even though there is yet no theoretical reason for doing so, \Eq{neutrino} may further invite to wonder what would be the Pauli pressure of a neutrino Fermi-gas, $P_{\nu}$, if it also had a density $\rho_{\nu} \sim \rho_{\Lambda}$. The result is $P_{\nu} \sim 1.5 \mathrm{x} 10^{-3} \rho_{\Lambda}$. If so, that would produce a QM repulsive contribution and a small correction to the exact equation of state, \Eq{darkenergy}, of the standard $\Lambda$-CDM model.\cite{wikiCMB} 

\section{Conclusions}\label{Conclusions}   

I developed an essential physical picture of dark energy based on most fundamental principles of quantum mechanics (QM) and general relativity (GR) theories. That derives from an ultimate mechanism of mass generation for initially massless gauge bosons in the standard model (SM) of quantum field theory (QFT) associated with quantum-mechanical confinement within a cosmological horizon.\cite{RescaPhoton} That is expressed in \Eq{fundamental}, which is in fact the only premise that I assume: all the rest follows from logic, well-established theory and elementary calculations.\cite{EinsteinPhoton}

Dark energy is intrinsically tied to de Sitter space, $dS_4$, and that has in turn a fundamentally invariant event horizon. I discussed how these matters are essential to my conjecture and model of dark energy and pressure, based on the standard $\Lambda$-CDM model of hot big bang cosmology. 

According to a possible solution of the flatness problem, I have considered a de Sitter universe with a Euclidean spatial geometry. That corresponds to flat slicing of $dS_4$ in FLRW geometry. I have also considered a de Sitter static metric, which has closed spatial sections. Based on further analysis,\cite{Weinberg, Moschella, Ibison, IbisonTwo} I have alternatively referred to properties of both flat and static slicings of $dS_4$.  

I conceive of dark energy as a Bose-Einstein condensate (BEC) of cosmologically massive photons and I have estimated fundamentally the binding energy per particle originating from the QM effectively attractive inter-particle potential in that BEC. Since massive photons may stand at rest in a de Sitter universe with flat spatial geometry, I solved the time-independent Schr\"{o}dinger equation for a non-relativistic attractive spherical-well potential self-confining at the de Sitter horizon. That provides the minimal critical potential depth that binds just one particle state at the top of that well. Combining that with the fundamental relativistic version of the uncertainty principle that cosmologically constrains the photon mass through its Compton wavelength at de Sitter horizon, \Eq{fundamental}, I obtained in \Eq{massvalue} a specific estimate of that mass, $m_g$, consistent with the dark energy-pressure relation of the standard flat $\Lambda$-CDM model.

My non-relativistic or `flat' estimate of $m_g$ is essentially confirmed by an independent relativistic or `curved' estimate of the QM zero-point energy of the BEC in de Sitter static metric with coordinate-time slicing. Remarkably, I can identify the dark energy density, $\rho_{\Lambda}$, with my zero-point energy density of the BEC, $\rho'_{0g}$, within a single order of magnitude. The Casimir electromagnetic energy density of the vacuum is of the order of $\rho_{CEM} \simeq \pi h c / l_P^4$ at a cut-off of the order of the Planck length.\cite{Ballentine} Thus the ratio of that $\rho_{CEM}$ to my $\rho'_{0g}$ in \Eq{zero-point} is of the order of $4.55 \mathrm{x} 10^{122}$, according to \Eq{Casimir-ratio}. On a logarithmic scale, that discrepancy is as large as anyone may get.\cite{Bull, Durrer, Steinhardt, Dadhich, BianchiRovelli, Weinberg, wikizero} 

Qualitative estimates that I provided in \Eq{rescaconstant} and \Eq{rescasecondconstant} are only meant to illustrate that numerical coefficients $C_g$ or $C_g'$ are of the order of $1$ or so. A far more precise evaluation of \Eq{fundamental} may only derive from QFT calculations of the $p_{\Lambda} / \rho_{\Lambda}$ equation of state and $g$-BEC properties in curved or de Sitter space-time.\cite{Steinhardt, Chavanis, WaldCurved, WaldHolland, Ford, Parker, BirrellCurved, Aldrovandi} 

%Then I estimated the gravitational pair-potential attraction between $m_g$ photons, $V_{12}$, and I found that its ratio to $V_{0c}$ is negligible.  

%Alternatively, I considered gravitational collapse for a uniform dark energy density, $\rho_{\Lambda}$, in an impending Schwarzschild geometry. Remarkably, that occurs at a Schwarzschild radius that equals $a_{\Lambda}$. The Newtonian gravitational force for a test particle of mass $m_g$ at that horizon turns out to coincide, but for a 1/2 factor, with the elementary non-relativistic QM formulation of $m_g$ binding.     

%An elementary physical picture of dark energy has thus emerged, as that of a BE condensate of self-attracting and ubiquitous $m_g$ photons fundamentally at rest in de Sitter universe. 

I further investigated statistical properties of equilibrium between the $g$-BEC phase and the ordinary vapor phase of kinetically energetic $m_g$-photons. I made comparisons with the Planck spectrum of the cosmic microwave background (CMB) and I found that corrections introduced by the tiny photon mass may be currently undetectable.

%Conceptually, however, a spin degeneracy factor of $g_S = 2$ applies to transverse massless photons, whereas $g_S = 3$ applies to massive photons, no matter how light. I discussed how this apparent discrepancy may be resolved. I proposed to regard the vapor phase as essentially composed of transverse massless photons, whereas in the $g$-BEC ground state, where the wavevector vanishes, both transverse and longitudinal $m_g$ photons may equally occur.

Remarkably, I considered a system of cosmological units, or `$g$-units,' that complements the fundamental system of Planck units in various ways, including uncertainty relations to which either set of units inherently correspond. Planck and $g$-units stand in a fundamental ratio given by \Eq{fundamentalratio}, spanning about 60 orders of magnitude. In the logarithmic middle of that range, the geometric mean of Planck and $g$-mass, \Eq{geometric}, or an equivalent but theory-independent \Eq{neutrino}, turn out to be tantalizingly close to current estimates of neutrino masses, almost within a single order of magnitude. I thus suggested that masses of the lightest known fermions may also be associated with both QM and GR fundamental constants, $\Lambda$, $G$, $c$ and $h$. 

Although based on entirely different physical grounds, there are common elements between my cosmological picture of dark energy and theories of ultra-light bosonic scalar and vector field dark matter,\cite{Calmet, ChavanisMass, Chavanis, Hwang, Hui, Hu, Li, Siemonsen, Tsukada, Morikawa, Nishiyama, Ferreira, LeeA} as well as theories of gravitational-vacuum and dark-energy stars.\cite{Mazur, Chapline, Laughlin, VisserGrava} Advanced techniques developed in those fields may also apply and advance my general formulation of dark energy theory. Other theories and observations may further probe my conjectures.\cite{Prat, Ruegg, Heeck, Reece, Nieto}

%Enable the following two lines to regenerate from refs.bib the .bbl file a la Francesco
%\bibliographystyle{ieeetr}
%\bibliography{refs.bib}

\begin{thebibliography}{10}
\bibitem{constant}
 The $A$ and $B$ constants in \Eq{MatterDensity} and \Eq{RadiationDensity}
  depend on the universe scale factor, $R(t)$. That may depend on an arbitrary
  constant, such as $\tilde{R}$ in \Eq{deSitterExpansion}.
\end{thebibliography}

%Incorporare il file .bbl nel file .tex e' un'operazione relativamente semplice su overleaf. L'unica cosa che devi fare e' estrarre il file .bbl dai log di overleaf main.tex, by clicking on the very last link at the right-bottom after recompiling and clicking on logfiles at the top, then download il file .bbl come txt, fare un copia ed incollare il file .bbl alla fine del file .tex sostituendo la parte "\bibliographystyle.... \bibliography.... " con l'intero file .bbl. Questo ti serve anche per uploadare il tuo file su arXiv.

\section{Appendix: Logotropic model and Cosmons}\label{Appendix}

After I completed my work and I submitted it for publication, its review by the Journal made me aware of some already published findings that are most relevant and possibly equivalent or deeply related to mine.\cite{ChavanisLogotropic, ChavanisCoreHalo, HarkoBohmerCosmon}

Specifically, the fundamental length and mass scales that I introduced in my \Eq{fundamental} and \Eq{mass} correspond to those of Eq. [C.9] and Footnote [8] of Ref. \onlinecite{ChavanisLogotropic}. The corresponding particle, called \textit{cosmon}, is associated with the current Hubble parameter, since in the \textit{logotropic model} the cosmological constant is ultimately time-dependent. Effectively, however, the current Hubble radius concurs with a fundamental cosmological constant corresponding to that of Einstein.\cite{ChavanisLogotropic, ChavanisCoreHalo} A cosmon mass scale has been interpreted as the smallest mass of bosonic particles of dark matter predicted by string theory,\cite{Axiverse} or as the \textit{upper bound} on the mass of gravitons.\cite{ChavanisLogotropic, ChavanisCoreHalo, DeFelice} 

On the other hand, I argued that a mass of the order of $10^{-32} eV/c^2$ must provide a \textit{lower bound} to the mass of any existing particle in the observable universe.\cite{RescaPhoton} That would rule out `cosminos' with an energy of the order of $10^{-96} eV$, which have been proposed as `quanta' of dark energy or the cosmological constant.\cite{HarkoBohmerCosmon} That paper asserts `that cosminos did not condense gravitationally, and hence the particles associated with dark energy fail to represent dark matter, which is in complete agreement with the present standard model of cosmology.' Whether conceptions of cosmons and cosminos developed in Ref. \onlinecite{HarkoBohmerCosmon} may be supported or not, perspectives of that paper are inspiring.   

In relation to models of dark matter halos, it has also been shown that the cosmon mass scale corresponds to the fundamental mass scale of BEC bosons: see Abstract, Eq. [158] and Eq. [I10] of Appendix [I-2] in Ref. \onlinecite{ChavanisCoreHalo}.

The neutrino mass scale that I introduce in my \Eq{geometric neutrino} and discuss in my \Secref{Neutrino} was also introduced in Ref. \onlinecite{ChavanisCoreHalo}. It is related to models of dark matter halos, consisting of fermions: see Abstract, Eq. [175] and Eq. [I4] of Appendix [I-1] in Ref. \onlinecite{ChavanisCoreHalo}.

Using various arguments and equations, Chavanis even relates in Eq. [37] and Eq. [38] of Ref. \onlinecite{ChavanisLogotropic} the electron mass and the fine structure constant to the GR and QM fundamental constants that I also consider in relation to the electron classical radius and its fermi scale, whose `coincidence' with $d_g$ I simply remark in the fourth paragraph of my \Secref{CMB}. 

So, the fundamental a-dimensional ratio that I consider in my \Eq{fundamentalratio} essentially coincides with Eq. [C.15] of Ref. \onlinecite{ChavanisLogotropic}. Likewise, my \Eq{Casimir-ratio} essentially coincides with Eq. [C.11] of Ref. \onlinecite{ChavanisLogotropic}. An estimate of the number of bosons in the observable universe is given in my \Eq{Archimedes} and correspondingly in Footnote [12] of Ref. \onlinecite{ChavanisLogotropic}.

Finally, the most tantalizing `coincidence' of the proton mass with my \Eq{proton} is also essentially found in Eq. [G.4] of Ref. \onlinecite{ChavanisLogotropic}. An equivalent relation is provided in Eq. [11] of Ref. \onlinecite{HarkoBohmerCosmon}. Further referring to the electron classical radius, B\"{o}hmer and Harko suggest a `small number hypothesis' as an extension of the famous `large number hypothesis' by Dirac. They propose that `the numerical equality between two very small quantities with a very similar physical meaning cannot be a simple coincidence.' I shall leave it at that, but Chavanis further provides incisive historical accounts of `curious coincidences' that intrigued famous scientists and surely many others time after time.\cite{ChavanisLogotropic} 

Evidently, the foundation, modelling and estimates of my work are altogether different and independent of those in Refs. \onlinecite{ChavanisLogotropic, ChavanisCoreHalo, HarkoBohmerCosmon} and possibly other works, quoted therein. What may then lie at the root of these remarkable `concurrences?' Well, at least some basic lines of reasoning seem relatively common throughout all these papers. But could there also be a deeper relation, or perhaps the actual coincidence, between the relatively `ordinary' $g$-photon that I propose at the root of it all and `cosmons' that have been variously envisioned in other work? Only much deeper and encompassing theory, observation and experiment may answer that fundamental question conclusively. But there will be an answer.

%Additional question: It is not quite clear where the attractive potentials (author-31) and (author-37) come from. They seem relatively ad hoc. Can the author give a more precise justification of these attractive potentials?

%I think that the paper of the author could be published after Refs. [a-c] have been properly quoted at the places where the author discusses similar results.

\section{Postscript: Fine Structure Constant}\label{Postscript}

The fine structure constant, $\alpha = e^2/\hbar c \simeq 1/137$, is an a-dimensional parameter that fundamentally characterizes quantum electrodynamics (QED). Using Planck units, I define $\alpha_P = e^2/h c \simeq 1/861$. Crude estimates of a numerical coefficient, $C_g$, that I provided in \Eq{rescaconstant} and in \Eq{rescasecondconstant} may reflect a far more fundamental relation, such as $C_g \sim \alpha_P$. If so, the photon mass in \Eq{mass} becomes
\begin{equation}\label{massfinestructure}
m_g c^2 \sim h c \alpha_P \sqrt{\Lambda} = e^2 / L_g \simeq 1.5 \mathrm{x} 10^{-41} MeV .
\end{equation}  
That of course does not mean that $L_g$ represents some sort of ``classical radius'' of the photon, which is neutral. Rather, \Eq{massfinestructure} indicates that the Coulomb potential of the electron, $e^2 /r$, which is carried by a massless photon to an infinite range for $r \rightarrow \infty$, reduces instead to a still vast, but cosmologically finite range, $L_g \equiv 1 /\sqrt{\Lambda} \simeq 10^{10} \mathrm{ly} \simeq 0.946 \mathrm{x} 10^{26} m$, when the electrostatic potential is carried by a massive $m_g$-photon. Thus, if confirmed, \Eq{massfinestructure} should demonstrate how profoundly a cosmological horizon and constant can affect and constrain quantum properties, propagation and interactions of QED and QCD elementary particles, including their mass generation through Higgs and St\"{u}ckelberg mechanisms. 

In fact, the counterpart of \Eq{massfinestructure} in the logotropic model is expressed in Eq. [32] and Eq. [37] of Ref. \onlinecite{ChavanisLogotropic}. With my units and definition of $\Lambda$, that Eq. [37] of Chavanis yields
\begin{equation}\label{massfinestructureChavanis}
m_e = 1.03 \alpha \bigg (\frac{\Lambda \hbar^4}{G^2} \bigg )^{(1/6)} \simeq 0.511MeV/c^2 
\end{equation}  
for the mass of the electron, $m_e$.

\end{document}